# A Modified Smoothed Particle Hydrodynamics Approach for Modelling Dynamic Contact Angle Hysteresis

Yanyao Bao, Ling Li, Luming Shen, Chengwang Lei, and Yixiang Gan *

School of Civil Engineering, The University of Sydney, NSW 2006, Australia

*Corresponding author's E-mail: yixiang.gan@sydney.edu.au

*Abstract*

Dynamic wetting plays an important role in the physics of multiphase flow, and has significant influence on many industrial and geotechnical applications. In this work, a modified smoothed particle hydrodynamics (SPH) model is employed to simulate surface tension, contact angle, and dynamic wetting effects. The wetting and dewetting phenomena are simulated in a capillary tube, where the liquid particles are raised or withdrawn by a shifting substrate. The SPH model is modified by introducing a newly-developed viscous force formulation at liquid-solid interface to reproduce the rate-dependent behaviour of moving contact line. Dynamic contact angle simulations with interfacial viscous force are conducted to verify the effectiveness and accuracy of this new formulation. In addition, the influence of interfacial viscous force with different magnitude on contact angle dynamics is examined by empirical power law correlations, and the derived constants suggest the dynamic contact angle changes monotonically with interfacial viscous force. The simulation results are consistent with the experimental observations and theoretical predictions, implying that the interfacial viscous force can be associated with slip length of flow and microscopic surface roughness. This work has demonstrated that the modified SPH model can successfully account for the rate-dependent effects of moving contact line, and can be used for realistic multiphase flow simulation under dynamic conditions.

# 1. Introduction

Complex interactions among multiple phases (e.g., gas, liquid, and solid) in porous media are of great significance in many industrial applications, such as groundwater treatment [1], oil recovery [2], carbon sequestration [3], and porous catalysts [4]. These processes usually involve dynamic wetting phenomena which are critical in describing the rate-dependent system properties, and can thus further shed light on optimising engineering solutions. Dynamic wettability is generally characterized by the dynamic contact angle. To access this property, previous studies focused on experimental characterisation [5-7], theoretical modelling [8,9], and various numerical approaches [10-12]. Numerical methods have demonstrated significant advantages in modelling multiphase flow [13,10,12,14], including time and cost efficiency, the wide range of length scales, and broadened material properties and loading conditions. However, to model dynamic wettability, most of existing numerical approaches are either far away from typical realistic time and length scales, e.g., using molecular dynamics (MD) [13], or established based on introducing additional artificial interfacial constrains, e.g., using smoothed particle dynamics (SPH) [12,14].

Dynamic contact angles can be measured experimentally at the liquid-solid-vapour triple-line region at different ranges of capillary number ($Ca$) [15,16], a non-dimensional parameter defined as the ratio of viscous forces to interfacial forces:

$$Ca = \frac{\eta v_t}{\gamma}, \tag{1}$$

where $v_t$ is triple-line region velocity, $\gamma$ is surface tension, and $\eta$ is viscosity of the fluid. The behaviours of dynamic contact angle associated with contact line motion of spreading liquids can be described by several theories, including hydrodynamic model [8], molecular-kinetic model [17], and their combination [18]. A number of empirical relationships based on these theoretical models for wetting have been discussed in the literature, all of which express the dynamic advancing (or receding) contact angle $\theta_d^a$ (or $\theta_d^r$) as a function of $Ca$ and the corresponding quasi static contact angle $\theta_s^a$ ($\theta_s^r$) during wetting (dewetting) processes. Elliot and Riddiford [19] found the dynamic contact angle is rate-independent when $Ca < 2 \times 10^{-7}$ for water with glass or polyethylene plate. Schwartz and Tejeda [5] identified that the dynamic contact angle is constant for $Ca < 2 \times 10^{-6}$. For $10^{-6} < Ca < 10^{-2}$, the dynamic contact angle changes monotonically with $Ca$, and the most commonly suggested relationship is [20-22]:

$$\left|\cos\theta_d^{a/r} - \cos\theta_s^{a/r}\right| = ACa^B, \tag{2}$$

where $A$ and $B$ are constants for advancing/receding cases. Based on the similar empirical correlation, Jiang et al. [6], Bracke et al. [7], and Seebergh and Berg [23] derived different constants for the proposed correlation, by introducing an additional term of $1/(1 + \cos\theta_s^{a/r})$ on the right hand side to minimize the deviations induced by fluid properties when comparing the results of different fluids. It has also been pointed out that the dependence of the contact angle on moving contact line velocity, i.e., the value of $B$, is enhanced by rough surfaces [23,5].

However, the complete understanding of dynamic contact angle associated with moving contact line is still an open topic due to the complex liquid-solid interactions and the fundamental roles of the triple-line region in the liquid spreading. Experimental investigations on dynamic contact angle could be strongly influenced by small-scale physical and chemical heterogeneities, impurities adsorbed on the solid surface, growth and dissolution of bubbles, etc [24]. Therefore, considering the above-mentioned limitations and availability of experimental conditions and facilities, the numerical approaches, including molecular dynamics (MD) [11,25,13], lattice-Boltzman methods (LBM) [10,26,27] and SPH [12,14], serve as powerful tools to study the contact angle dynamics and the fundamental mechanisms therein. At the micro-scale, Koplik et al. [11] identified rate-dependent behaviour for dynamic receding angle with MD simulation in an immiscible two-fluid system. Lukyanov and Likhtman [13] applied MD simulation to explain the behaviours of dynamic contact angle from the perspective of force distribution and friction law. Nevertheless, there are restrictions in simulation time and length scales due to high computing power required for MD simulations. At the meso-scale, researchers also proposed and verified the power law correlation between $Ca$ and dynamic contact angle with multiphase capillary flow using LBM and SPH method, and the outcomes consist with experimental results and theoretical predictions [10,26,27,12,14]. To deal with moving boundaries problem, LBM requires additional algorithm which may lose the accuracy of the standard scheme [28]. For the abovementioned SPH models that successfully simulated the dynamic contact angle, additional terms should be imposed, such as contact line force formulation [12] at the triple-line region and Young–Laplace boundary condition [14] at the fluid-fluid interface, all of which require explicit modelling of all phases in the pore space, and thus dramatically increase the computational cost.

In this paper, we propose a modified SPH model to study the contact angle dynamics. This SPH model adopts additional liquid-liquid and liquid-solid interaction forces to generate the surface tension and wetting effects [48]. Moreover, to replicate realistic dynamic behaviour of liquid-solid interactions, an interfacial viscous force is introduced to capture the viscous shearing, which was omitted in the original SPH formulation. This newly-introduced interfacial force is formulated based on the physically measurable quantity, i.e., the slip length, and demonstrated to correlate with surface chemistry and roughness conditions. Advancing and receding contact angles are simulated in a capillary tube under various contact line speeds. By analysing the correlations between the contact line velocity and contact angle, empirical power law correlations with various constants are obtained. Furthermore, parametric studies have been conducted to demonstrate that these predicted dynamic behaviours correlate to the proposed interfacial viscous force model.

## 2. SPH Model

### 2.1. Governing equations

The motion of an incompressible fluid is governed by Navier–Stokes equations which can be written in Lagrangian form as follows:

$$\frac{d\rho}{dt} = -\rho \nabla \cdot \boldsymbol{v}, \qquad (3)$$

$$\frac{d\boldsymbol{v}}{dt} = -\frac{\nabla P}{\rho} + \mu \frac{\nabla^2 \boldsymbol{v}}{\rho} + \boldsymbol{F}, \qquad (4)$$

where $\rho$ is fluid density, $\boldsymbol{v}$ is flow velocity, $P$ is pressure, $\mu \frac{\nabla^2 \boldsymbol{v}}{\rho}$ is the viscous term, and $\boldsymbol{F}$ corresponds to total volumetric force acting on unit mass.

We approximate the incompressibility of fluids using the weakly-compressible approach by applying the Equation of State (EOS) with the form of [29]:

$$P = \frac{c^2 \rho_0}{\chi}\left[\left(\frac{\rho}{\rho_0}\right)^\chi - 1\right], \qquad (5)$$

where $c$ is artificial speed of sound, $\rho_0$ is reference density and $\chi$ is empirical parameter. In this work, we choose the artificial speed of sound as $c > 10 V_{max}$ along with $\chi = 7$ [30,31], where $V_{max}$ is the expected maximum particle velocity.

### 2.2. The SPH equations

The SPH method is based on the idea that a continuous field $A(\boldsymbol{r}_i)$ at position $\boldsymbol{r}_i$ can be smoothed by a convolution integral with smoothing function, $W(\boldsymbol{r}_i - \boldsymbol{r}_j, h)$, allowing the value of any function to be obtained at a given point with neighbouring particles [32]:

$$A(\boldsymbol{r}_i) = \sum_{j=1}^{N} \frac{m_j}{\rho_j} A(\boldsymbol{r}_j) W(\boldsymbol{r}_i - \boldsymbol{r}_j, h), \qquad (6)$$

where the summation is over $N$ neighbouring particles around the specified particle $\boldsymbol{r}_i$, and $m_j$, $\boldsymbol{r}_j$, $\rho_j$ are the mass, position and the local density of particle $j$ respectively, $W(\boldsymbol{r}_i - \boldsymbol{r}_j, h)$ is the smoothing function with $h$ as its smoothing length. The Gaussian kernel is adopted for the proposed SPH model considering both calculation accuracy and computational efficiency [33]:

$$W(\boldsymbol{r}, h) = \frac{1}{h\sqrt{\pi}} e^{-\left(\frac{r^2}{h^2}\right)}. \tag{7}$$

The density of a liquid particle $i$ is evaluated by:

$$\rho_i = \sum_j m_j W(\boldsymbol{r}_i - \boldsymbol{r}_j, h). \tag{8}$$

The pressure driven part of the momentum equation is implemented following the approach proposed by Monaghan [32], the pressure gradient is symmetrized by rewriting $\nabla P/\rho$ in Eq. (4) to ensure momentum conservation:

$$\nabla P_i = \rho_i \sum_j m_j \left[\frac{P_i}{\rho_i^2} + \frac{P_j}{\rho_j^2}\right] \nabla W(\boldsymbol{r}_i - \boldsymbol{r}_j, h). \tag{9}$$

A Monaghan style artificial viscosity model [32] is applied to stabilize the numerical algorithm. The artificial viscosity is obtained by writing momentum equation as the following form:

$$\mu \frac{\nabla^2 \boldsymbol{v}_i}{\rho_i} = -\sum_j m_j \Pi_{ab} \nabla W(\boldsymbol{r}_i - \boldsymbol{r}_j, h), \tag{10}$$

where $\Pi_{ab}$ is given by:

$$\Pi_{ij} = \begin{cases} \dfrac{-\alpha \bar{c}_{ij} \mu_{ij} + \beta \mu_{ij}^2}{\bar{\rho}_{ij}} & \boldsymbol{v}_{ij} \cdot \boldsymbol{r}_{ij} < 0; \\ 0 & \boldsymbol{v}_{ij} \cdot \boldsymbol{r}_{ij} > 0; \end{cases} \tag{11}$$

and

$$\mu_{ij} = \frac{h \boldsymbol{v}_{ij} \cdot \boldsymbol{r}_{ij}}{\boldsymbol{r}_{ij}^2 + 0.01 h^2}, \tag{12}$$

where $\alpha$ and $\beta$ are constants, $\bar{c}_{ij} = (c_i + c_j)/2$ and $\bar{\rho}_{ij} = (\rho_i + \rho_j)/2$ are the values of the sound speed and the density averaged between particles $i$ and $j$. Monaghan [32] suggested the values for $\alpha$ and $\beta$ are 1 and 2 respectively for best results. In this work, we adopt the $\alpha$ value while setting the value of $\beta$ to 0 as the motion of fluid flow in the simulation is relatively slow (ranging from 0.0002 to 20 mm/s). It should be noted that the implementation of this artificial viscosity could lead to unphysically high shear viscous forces because SPH simulations are stabilized by the physical viscosity of a fluid, and it is challenging to simulate low viscous flows [34]. Therefore, our simulations will be carried out with this high viscous setting.

For the weakly-compressible SPH formulation, the time step $\Delta t$ in this work follows the CFL-condition based on: (1) the maximum artificial sound speed and the maximum flow speed; (2) the magnitude of particle acceleration $f_i$; and (3) the viscous condition [35].

## 2.3. Implementation of inter-particle force model in SPH

In this work, an inter-particle force formulation proposed by Li et al. [36] is applied to reproduce the surface tension and wetting effects. The pair potential energy $U(r)$ and inter-particle force $\boldsymbol{F}_{ij}^{inter}$ are given as:

$$U(r) = \alpha_{ij}[CW^e(r,h_1) - DW^e(r,h_2)], \text{ where } h_1, h_2 \leq 0.5h, \qquad (13)$$

$$\boldsymbol{F}_{ij}^{inter} = -\frac{dU(r)}{dr}\frac{\boldsymbol{r}_i - \boldsymbol{r}_j}{\|\boldsymbol{r}_i - \boldsymbol{r}_j\|}, i \neq j, \qquad (14)$$

where $\alpha_{ij}$ is inter-particle force strength parameter, $C$ and $D$ are constants, $W^e$ is the cubic spline function to construct the potential energy function with the form of:

$$W^e(r,h) \begin{cases} 1 - \frac{3}{2}\left(\frac{r}{h}\right)^2 + \frac{3}{4}\left(\frac{r}{h}\right)^3 & 0 < \frac{r}{h} \leq 1 \\ \frac{1}{4}\left(2 - \frac{r}{h}\right)^3 & 1 < \frac{r}{h} \leq 2 \\ 0 & otherwise \end{cases}. \qquad (15)$$

This kernel function $W^e$ is normalized by $\frac{2}{3h}, \frac{10}{7\pi h^2}, \frac{1}{\pi h^3}$ in 1D, 2D and 3D space respectively. The inter-particle force in Eq. (14) is imposed on both liquid and solid particles to generate multiphase interactions between different phases. The parameter $\alpha_{ij}$ is replaced by $\alpha_1$ (as the liquid-liquid interaction force parameter) when particle $j$ is a liquid particle and by $\alpha_2$ (liquid-solid interaction force parameter) when particle $j$ is a solid particle. Note that particle $i$ is always a liquid particle. By assigning different value for $\alpha_1$ and $\alpha_2$, different surface tension of free surface and static contact angle on solid surface can be reproduced in the simulation. In addition, to prevent the unphysically penetration of liquid particles into the solid particles, additional repulsive boundary forces are implemented similar to those implemented in Monaghan and Kajtar [37].

## 2.4. Modified liquid-solid interface model

In most SPH simulations, the boundaries of rigid bodies have been prescribed with different mechanisms, such as ghost particles [38], normalising conditions [39], and boundary particle forces [37], all of which are only appropriate for perfectly smooth boundaries [40]. Considering the particle-particle interaction force at liquid-solid interface, the summation of the short range repulsive and longer range attractive forces acting on the liquid particle is fluctuating around zero on the tangential direction, which effectively makes the interface frictionless and rate-independent, see Figure 1(b). Therefore, the rate-dependent behaviour of moving contact line is hardly achieved in this circumstance.

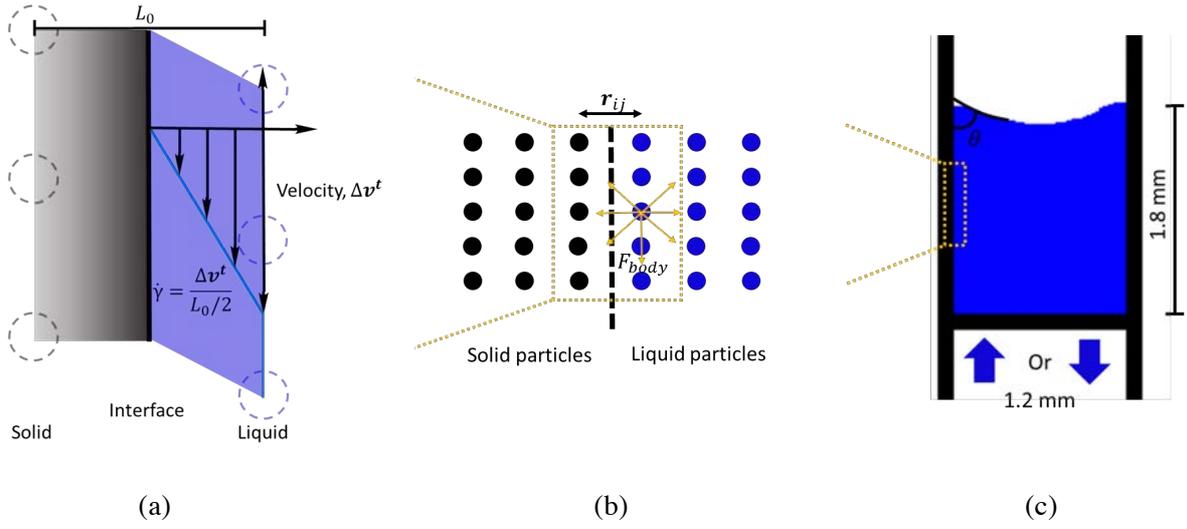

Fig. 1 Schematics of the modified SPH approach (not to scale): (a) Shearing between solid and liquid particles at the liquid-solid interface; (b) Force balance of a liquid particle; (c) Geometry of capillary tube for dynamic contact angle simulation.

There are several approaches in SPH to prescribe dynamic contact angle at the contact line, e.g., using Young–Laplace boundary condition at the fluid–fluid interface [14], and introducing a contact line force model [12]. However, the above approaches require explicit modelling of the geometries and boundaries of all phases in the pore space. Thus the computational cost is dramatically increased. In this work, we introduce a new algorithm which imposes a viscous force $\boldsymbol{F}_{ij}^{vis}$ on the liquid particles at liquid-solid interface to reproduce the rate-dependent behaviour of moving contact line. For an ideal no-slip condition, we can express the tangential force between solid and liquid as

$$\boldsymbol{F}_{ij}^{vis} = -S \cdot \tau \cdot \boldsymbol{t} = -L_0 \cdot \eta \cdot \dot{\gamma} \cdot \boldsymbol{t} = -L_0 \cdot \eta \cdot \frac{\Delta \boldsymbol{v}^t}{L_0/2}, \qquad (16)$$

where $S$ is liquid-solid contact area, and in two-dimensional case the contact area becomes the particle spacing $L_0$, $\tau$ is shear stress, $\eta$ is the viscosity of bulk fluid, $\dot{\gamma}$ is shear rate, $\boldsymbol{t}$ is the tangential unit vector, and $\Delta \boldsymbol{v}^t$ is the relative tangential velocity between liquid and solid particles. The thickness of shearing equals to half of the interface spacing $L_0$, as shown in Figure 1(a). This formulation is based on assumptions of the laminar flow with a constant shear rate and no-slip boundary condition at the liquid-solid interface.

Experimental studies have reported the slip length ranging from 10 nm to 10 μm with various surfaces and liquids [41-44]. However, compared with the predictions under no-slip boundary conditions and bulk fluid viscosity, the average boundary fluid velocity and slip length of pressure-driven flow can increase dramatically on certain surfaces, such as superhydrophobic and rough surfaces [45-48]. For example, the slip length can reach hundreds of nanometres for hexadecane flowing over a bare sapphire surface [45], up to 400 um on hydrophobic micro-macro structures [48], and even larger than 1 mm for fluid flow through the aligned carbon-nanotube membrane [47]. In addition, MD simulations also suggest the slip length may increase for fluids at triple-line region [11].

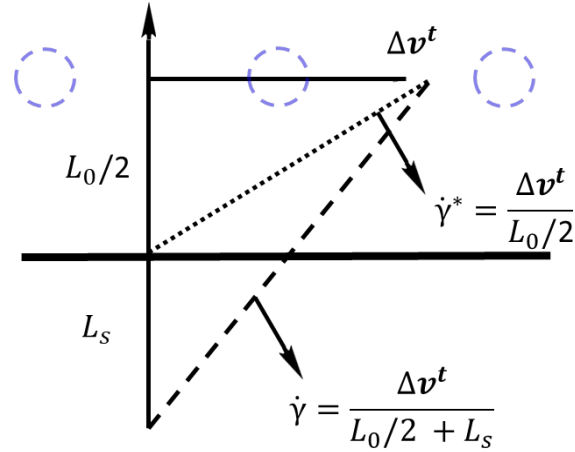

Fig. 2 Velocity profiles and shear rate for: slip condition (dashed line), and the equivalent no-slip condition (dotted line).

To take the presence of the slip length into account, the interfacial force formulation in Eq. (16) needs to be modified. Figure 2 illustrates the equivalent model for effective shear rate $\dot{\gamma}^*$ considering the slip boundary condition with a slip length of $L_s$. The equivalent model should have the same tangential force for the actual case with a slip condition, as:

$$\boldsymbol{F}_{ij}^{vis} = -L_0 \cdot \eta^* \cdot \frac{\Delta \boldsymbol{v}^t}{L_0/2} = -L_0 \cdot \eta \cdot \frac{\Delta \boldsymbol{v}^t}{L_0/2 + L_s}, \tag{17}$$

where $\eta^*$ is the equivalent viscosity, or apparent viscosity, at the interface. Thus, we have:

$$\eta^* = \eta \cdot \frac{L_0}{L_0 + 2L_s} . \tag{18}$$

Therefore, the bulk viscosity $\eta$ at interface is replaced by a smaller value $\eta^*$ ($L_s > 0$) to reproduce the actual shear profile at the liquid-solid interface. Finally, we have interfacial viscous force $\boldsymbol{F}_{ij}^{vis}$ added in the inter-particle force formulation with the following form:

$$\boldsymbol{F}_{ij}^{vis} = \begin{cases} -L_0 \cdot \eta^* \cdot \frac{\Delta \boldsymbol{v}^t}{L_0/2}, & r_{ij} \leq L_0 \\ 0, & r_{ij} > L_0 \end{cases} . \tag{19}$$

Mesh sensitivity studies have been conducted and the results suggest that the use of Eqs. (18) and (19) is independent of particle spacing $L_0$. Modified with the inter-particle interaction force, $\boldsymbol{F}_{ij}^{inter}$, and interfacial viscous force, $\boldsymbol{F}_{ij}^{vis}$, the SPH discretization of the governing equation now can be read as follows:

$$\frac{d\boldsymbol{v}_i}{dt} = -\sum_j m_j \left[\frac{P_i}{\rho_i^2} + \frac{P_j}{\rho_j^2} + \Pi_{ab}\right] \nabla W(\boldsymbol{r}_i - \boldsymbol{r}_j, h) + \boldsymbol{g} + \sum_j \frac{\boldsymbol{F}_{ij}^{inter}}{m_i} + \sum_j \frac{\boldsymbol{F}_{ij}^{vis}}{m_i}. \qquad (20)$$

This SPH model is implemented in an open source framework PySPH [49], and the parameters used in this work are listed in Table 1 unless otherwise mentioned.

**Table 1** Parameters for dynamic contact angle simulations.

| Parameters | Symbols | Value |
| --- | --- | --- |
| Density (kg/m$^3$) | $\rho_0$ | 1000.0 |
| Gravity (m/s$^2$) | $g$ | 9.8 |
| Viscosity (Pa·s)$^a$ | $\eta$ | 0.013, 0.04, 0.185 |
| Interfacial viscosity (Pa·s) | $\eta^*$ | 0.0006 ~ 0.009 |
| Slip length (μm)$^a$ | $L_s$ | 100 ~ 1000 |
| Surface tension (N/m)$^a$ | $\gamma$ | 0.28 |
| Particle mass (kg) | $m_0$ | $6.25 \times 10^{-7}$ |
| Particle spacing (m) | $L_0$ | $2.5 \times 10^{-5}$ |
| Smoothing length (m) | $h$ | $7.5 \times 10^{-5}$ |
| Artificial sound speed (m/s) | $c$ | 3.0 |
| Time step (s) | $\Delta t$ | $4.15 \times 10^{-7}$ |
| Liquid-liquid interaction force parameter (J) | $\alpha_1$ | $5.42 \times 10^{-4}$ |
| Liquid-solid interaction force parameter (J) | $\alpha_2$ | $5.69 \times 10^{-4}$ |

$^a$These material properties are the results of other input parameters.
$^b$The values here correspond to the interfacial viscosity used in the simulations.

## 3. Results and discussion

### 3.1. Identification of model parameters

To examine the surface tension scheme used in this study, the shape evolution of a 2D droplet with zero gravity in vacuum is presented. Figure 3(a) shows the shape transformation of a droplet from square to circle due to the surface tension effect. The surface tension $\gamma$ can be calculated by the Young-Laplace equation which relates the difference of inside and outside pressure of the droplet $\Delta P$ and its radius $R$:

$$\gamma \left(\frac{1}{R_1} + \frac{1}{R_2}\right) = \Delta P, \tag{21}$$

where $R_1$ and $R_2$ are the principal radii of droplet curvature. Since the 2D droplet is simulated in vacuum, $\Delta P$ equals to the pressure inside the droplet and the mean radius equals to the droplet radius $R$. In this work, the total pressure of the droplet is the sum of pressure calculated from EOS plus the contribution by inter-particle force $F_{ij}^{inter}$. Due to the boundary deficiency [50], we exclude the edge of droplet and use bulk region for pressure measurement. Seven tests are conducted with different value of $\alpha_1$ ranging from $2.0 \times 10^{-5}$ to $5.5 \times 10^{-4}$ J, and the results suggest a linear relationship between $\alpha_1$ and $\gamma$, see Figure 3(b).

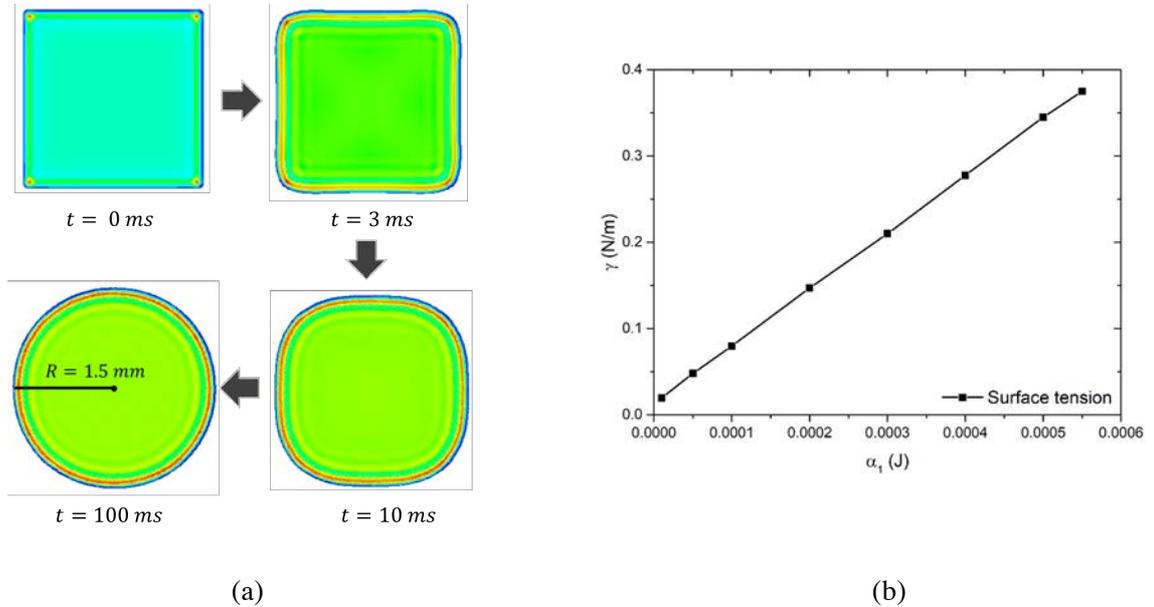

(a)  (b)

Fig. 3 (a) Evolution of droplet shape; (b) Surface tension with different value of $\alpha_1$.

The liquid-solid interaction strength parameter $\alpha_2$ is calibrated by simulating six different equilibrium contact angles of droplet over a flat substrate, with a thickness of 0.1 mm, as shown in Figure 4(a). The droplet has a volume of 2.5 mm³, and it is slowly brought into contact with the flat surface under gravity force. After the droplet reaches equilibrium (approximately 0.25 s), the curvature of droplet is fitted for contact angle measurement. Figure 4(a) and (b) show that different wetting behaviour from hydrophobicity to hydrophilicity can be simulated by adjusting the liquid-solid interaction strength parameter $\alpha_2$.

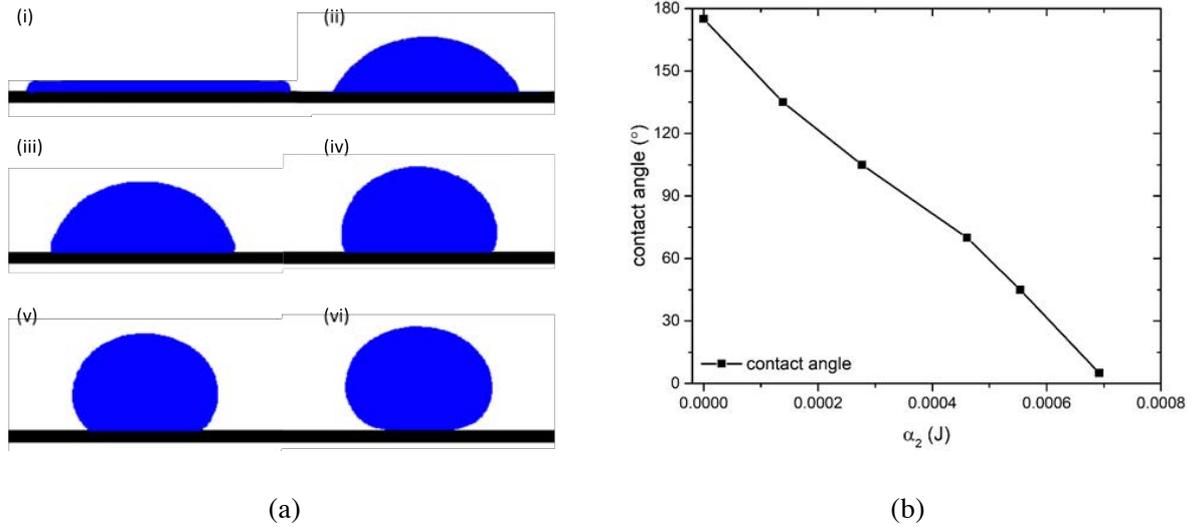

(a)            (b)

Fig. 4 (a) Static contact angles for different $\alpha_2$; (b) Relation between static contact angle and $\alpha_2$.

## 3.2. Dynamic contact angles

To simulate the moving contact line and dynamic contact angle, a 2D capillary tube with a shifting substrate is modelled in vacuum with the following geometry and parameters. The size of the capillary tube is 4 mm × 1.36 mm containing fluid with domain of 1.8 mm × 1.2mm, as shown in Figure 1(c). The surface tension and bulk viscosity of the simulated fluid is fixed unless mentioned otherwise, and $Ca$ is only controlled by the triple line velocity, $v_t$, in this work. The contact angle $\theta$ is considered as the included angle by the solid boundary and the tangent of the fluid curvature close to the liquid-solid interface. The first three layers of liquid particles adjacent to the solid boundary are excluded for contact angle measurement as these particles frequently reconstruct due to the liquid-solid interactions and boundary repulsive force, resulting in an unstable triple line region.

Prior to the simulations of different substrate moving speed, the capillary tube is under the same stable configuration in terms of fluid curvature and static contact angle. In all simulation cases, the bottom substrate moves along the *y*-direction for a distance of 1.2 mm to raise or withdraw the fluid with various velocities to generate a moving contact line. Figure 5 shows snapshots of fluid movement and curvature at different simulation time steps. The advancing angle occurs when the fluid is pushed up, and receding angle is formed when the substrate withdraws the fluid. At any given time, the resulting contact angle will be recorded against the instantaneous contact line speed.

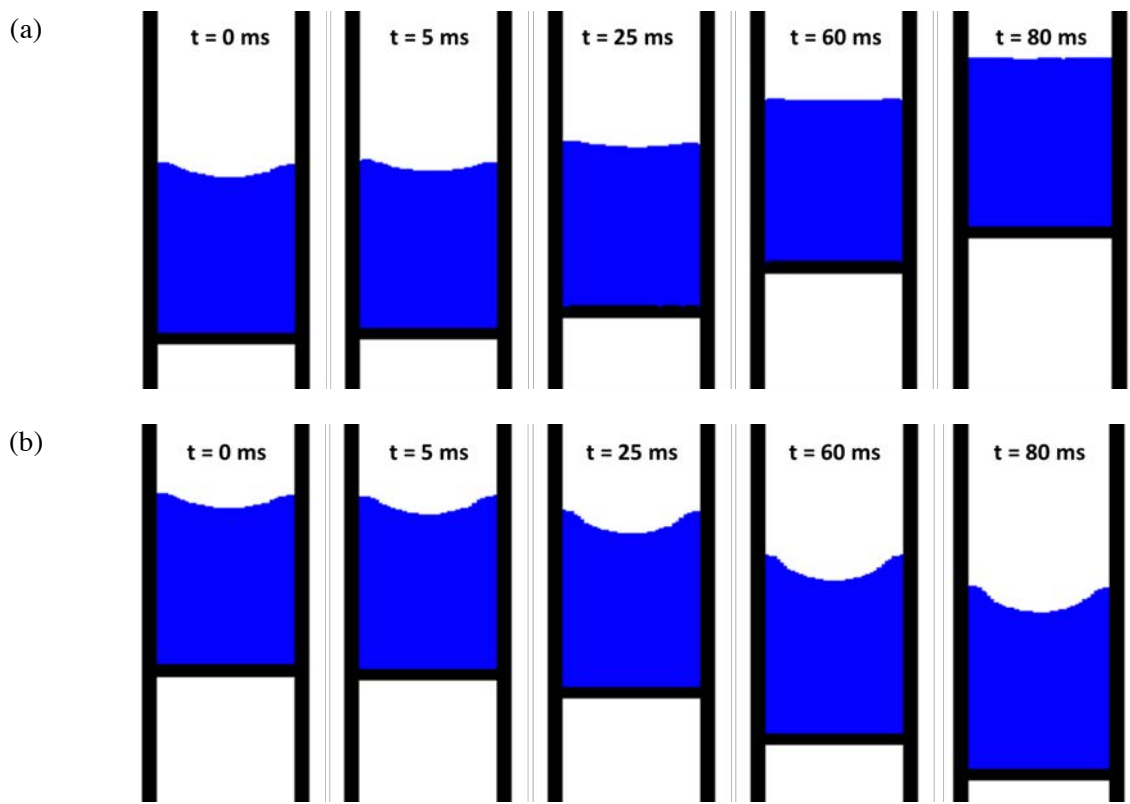

Fig. 5 Snapshots of dynamic contact angle simulation at various time while the moving speed of substrate equals 10 mm/s: (a) advancing case; (b) receding case.

In Figure 5, five snapshots of capillary tube with substrate moving velocity of 10 mm/s at simulation time 0 ms, 5 ms, 25 ms, 60 ms and 80 ms are presented. At the beginning of the simulation, due to the weakly compressibility and low sound speed of the simulated fluid, it takes hundreds of time-steps for substrate transferring its speed to the whole liquid particles. Subsequently, the fluid curvature starts to change and the advancing (or receding) angles keep increasing (or decreasing) from simulation time 0 ms to 60 ms. After 60 ms, the fluid curvature becomes stable and the contact angle measurement is conducted afterwards. It should be noted that the velocity of substrate does not necessarily equal to the velocity of triple-line region, and the position and velocity of liquid particles are updated according to the smoothing function at each time-step, so the fitted contact angle and $Ca$ will be varying during the simulation. Therefore, when the curvature becomes stable, i.e., after simulation time 60 ms in this case, hundreds sets of contact line moving velocity and dynamic contact angle are recorded for further data processing and analyses.

To examine the effect of the proposed interfacial viscous force, we conducted two dynamic contact angle simulations, implemented with and without the term $\boldsymbol{F}_i^{vis}$ (more specifically, $\eta^* = 0.003$ Pa·s and $\eta^* = 0.0$ Pa·s, respectively), and the results are compared in terms of dynamic contact angle in Figure 6. Both cases are simulated with 12 sets of substrate moving speed ranging from 0.002 mm/s to 20 mm/s. The resulted $Ca$ covers from $10^{-6}$ to $10^{-1}$.

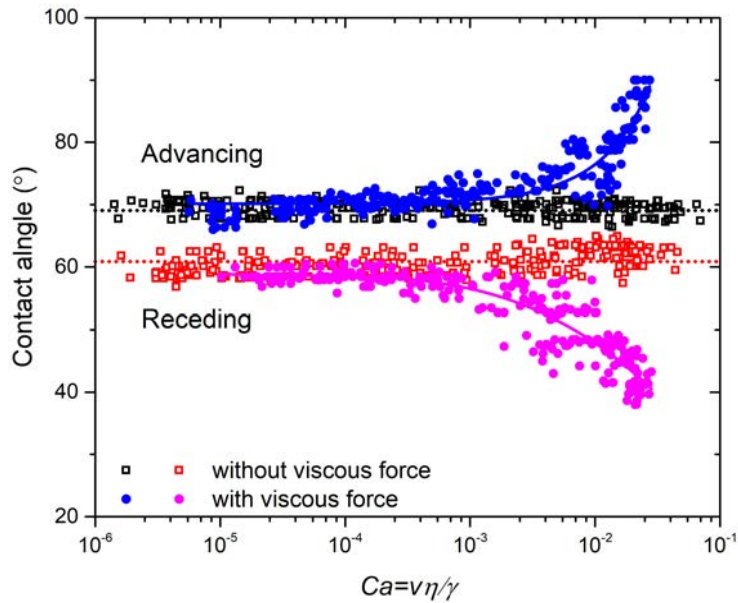

Fig. 6 Scattered plot of dynamic contact angle simulation with and without interfacial viscous force, $\boldsymbol{F}_i^{vis}$. Blue and pink circles represent advancing and receding cases with $\boldsymbol{F}_i^{vis}$, respectively; Black and red squares: advancing and receding cases without $\boldsymbol{F}_i^{vis}$. Here, trend lines are guide for eyes.

For $Ca < 10^{-4}$, the dynamic contact angle simulation results are independent of $\boldsymbol{F}_i^{vis}$ where the dynamic advancing and receding angle is around 70° and 60°, respectively. For $10^{-4} < Ca < 10^{-3}$, slight difference can be observed from the results. For cases implemented with $\boldsymbol{F}_i^{vis}$, the dynamic advancing angle starts to increase and the receding angle starts to decrease, while the dynamic contact angle in the case without $\boldsymbol{F}_i^{vis}$ remains unchanged. The major difference occurs in the large $Ca$ regime ($Ca > 10^{-3}$): With the increase of $Ca$, the dynamic advancing/receding contact angle keeps almost constant in the case without $\boldsymbol{F}_i^{vis}$, which suggests there is no rate-dependent behaviour of dynamic contact angle. While for cases implemented with $F_i^{vis}$, the dynamic advancing angle increases from 70° to around 90° and the dynamic receding angle decreases from 60° to less than 40° when $Ca$ approaches 0.02.

The simulation results suggest that the interfacial viscous force formulation is the key for reproducing dynamic contact angle. According to Eq. (19), the interfacial viscous force $F_i^{vis}$ at liquid-solid interface region is increased with the velocity of liquid particles. Therefore, $F_i^{vis}$ formulation will result in different influence on the motion of moving contact line depending on the magnitude of the triple velocity, $v_t$. In the case where $v_t$ is relatively small ($Ca < 10^{-4}$), the magnitude of interfacial viscous force $F_i^{vis}$ is also negligible and the contact angle hysteresis is hardly observed within small $Ca$ regime. When $v_t$ becomes larger, $F_i^{vis}$ starts to influence the behaviour of moving contact line. In such a circumstance, the relative motion of contact line is restricted and slowed down along the flow direction. Meanwhile, the bulk fluid in the middle region of capillary tube is not affected by the interfacial viscous force. Hence this part of fluid moves quicker than the fluid at triple-line region, which creates a larger (or smaller) contact angle for advancing (or receding) case. In summary, with the adoption of the newly-introduced interfacial viscous force formulation, the proposed SPH model can successfully simulate the dynamic contact angle.

In the following, empirical power law correlations are used to examine the results, similar to the format of the "universal function" [23]. The above simulation results with the presence of $F_i^{vis}$ are replotted in Figure 7 using smoothed density histogram plot. Note that the legend bars represent the relative probability densities, e.g., '1.0' refers to the highest density. Then, the power law correlation of Eq. (2) is used to fit the results and the constants are derived. From previous studies, we know that the exponent $B$ ranges from 0.2 to 1.0 based on various experimental findings and theoretical prediction [20]. In our simulation results with $\eta^* = 0.003$ Pa·s, $B$ values of 0.531 and 0.406 are derived for advancing and receding cases, respectively. Later in Section 3.3, we will further examine the dependency of $\eta^*$ for both the exponent and amplitude constants. Here, the quasi-static advancing and receding contact angle ($\theta_s^a = 69.53°$ and $\theta_s^r = 59.82°$) are obtained by raising or lowering the bottom substrate at extremely slow velocity of $2\times10^{-5}$ mm/s.

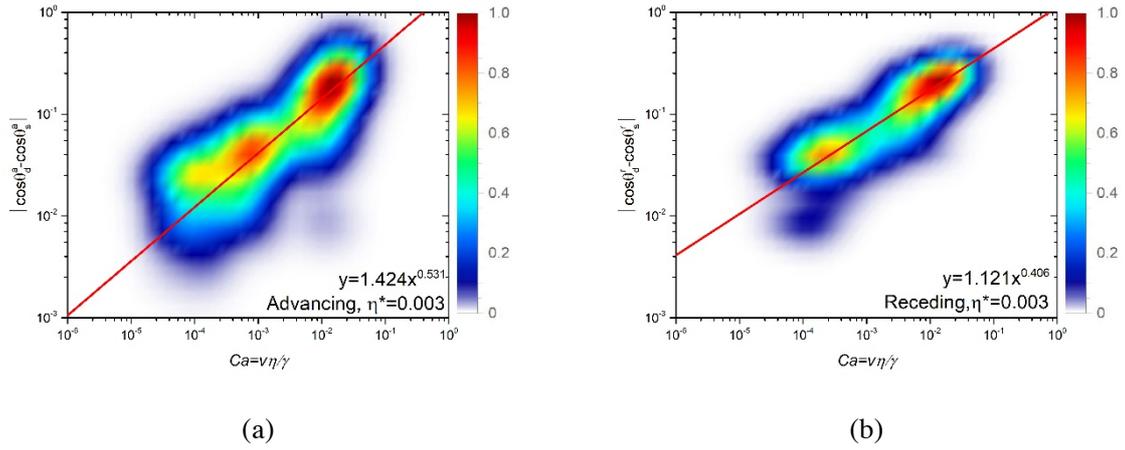

Fig. 7 Density histogram plots and power law fitting for dynamic (a) advancing and (b) receding contact angle results, x and y stands for $Ca$ and $\left|\cos\theta_d^{a/r} - \cos\theta_s^{a/r}\right|$, respectively.

### 3.3. Parametric study of $\eta^*$

In this section, we present the parametric study on the interfacial viscous force $\boldsymbol{F}_i^{vis}$ formulation, i.e., $\eta^*$. In order to study the correlation between interfacial viscous force parameter $\eta^*$ and resulting dynamic contact angle, six sets of simulations with $\eta^*$ ranging from 0, 0.0006, 0.0015, 0.003, 0.006 and 0.009 Pa·s are conducted. The receding case with $\eta^* = 0.009$ Pa·s is not shown due to the strong adhesion between liquid and solid phase, which will be explained later in this section. Each set of simulations are conducted with different substrate speeds, resulting $Ca$ covering from $10^{-6}$ to $10^{-2}$.

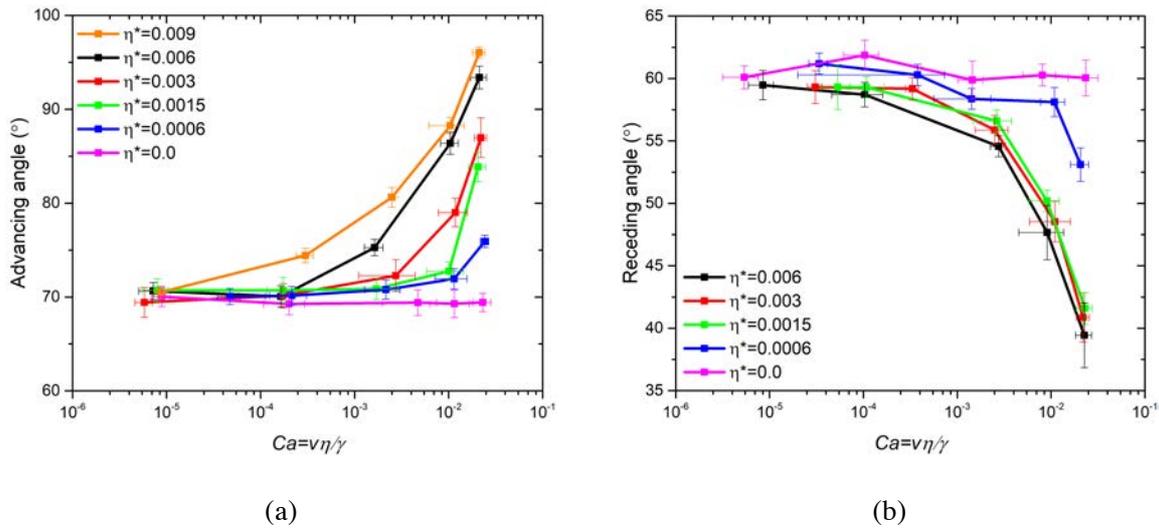

Fig. 8 Dynamic contact angle with different magnitude of interfacial viscous force for (a) advancing cases; (b) receding cases.

We selected the simulations with bottom substrate speeds of 20 mm/s, 10 mm/s, 2 mm/s, 0.2 mm/s, 0.02 mm/s and 0.002 mm/s to study the correlation between dynamic contact angle and $Ca$ with different $\eta^*$. In Figure 8, it can be seen that larger $\eta^*$ results in a larger advancing angle and smaller receding angle especially at higher $Ca$ region, which means the contact angle hysteresis is enhanced with the increase of $\eta^*$. As $Ca$ is getting smaller, the dynamic advancing and receding angle converge to around 70° and 60°, respectively.

The enhancement of dynamic contact angle on larger $\eta^*$ can be interpreted from the perspective of slip length and surface roughness, and this phenomenon is in good agreement with experimental results. Studies have suggested that the slip length is reduced by surface roughness [51], and rough surfaces can enhance the contact angle hysteresis [52]. According to Eq. (18), the value of $\eta^*$ is inversely proportional to the slip length $L_s$ in the model. Therefore, the selection of relatively large value $\eta^*$ refers to a small $L_s$, representing a rough surface, and such a surface is observed to have relatively large contact angle hysteresis, consistent with the experimental observations.

Nevertheless, in the receding case when $\eta^*$ is larger than 0.0015 Pa·s, the dynamic angles in high $Ca$ regime are less dependent on the actual values of $\eta^*$. The liquid-solid particle interactions with strong $\boldsymbol{F}_i^{vis}$ is considered to be responsible for this phenomenon. In the relative large $\boldsymbol{F}_i^{vis}$ scenarios, a thin water film is formed and attached on the solid surface and gravity force can hardly drive it downwards, see Figure 9(a). In such case, the dynamic receding angle is perceived as the intersection between the water film and fluid curvature, and the angles converge for $\eta^*$=0.0015, 0.003 and 0.006 Pa·s. When the moving contact line velocity is smaller, or $\boldsymbol{F}_i^{vis}$ is less significant, there will be no water film formed on the solid surface as shown in Figure 9(b).

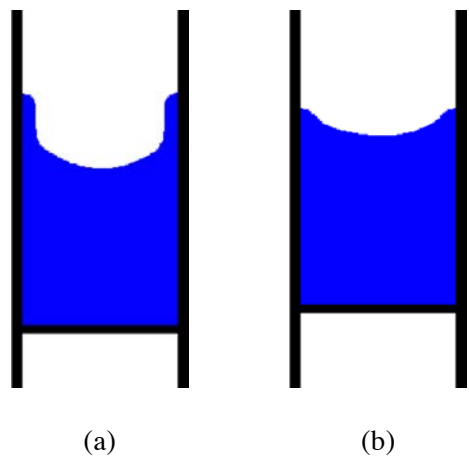

(a)      (b)

Fig. 9 Dynamic receding angle simulations with $\eta^*$=0.006 Pa·s, (a) presence of thin water film with the substrate moving at 20 mm/s; (b) without thin water film with the substrate moving at 2 mm/s.

To obtain a further understanding of how the parameter $\eta^*$ influences the dynamic contact angle, the simulation results with five different $\eta^*$ values are fitted with Eq. (2), and amplitude $A$ and exponent $B$ are derived correspondingly. In Figure 10, it is observed that $A$ and $B$ in the power law fitting change with the value of $\eta^*$ (only limited cases are shown due to the space). For $\eta^* \geq 0.0015$ Pa·s, the slope is quite obvious, the value of $\left|\cos\theta_d^{a/r} - \cos\theta_s^{a/r}\right|$ increases with the increase of $Ca$. For the case $\eta^* = 0$, the slope is 0 as expected since there is no rate-dependent behaviour of contact angle.

Furthermore, we compared our numerical predictions using $\eta^*$ = 0.009, 0.006 and 0.003 Pa·s with the empirical power law correlations proposed in [6,7] and experimental data extracted from [53] that including various combination of liquid and solid materials. Since all these previous studies focused on the dynamic advancing cases for various types of liquid, we used here the additional term $(1 + \cos\theta_s^a)$ to unify the numerical and experimental data, as shown in Figure 11. Excellent agreement is observed between simulation results with $\eta^*$ = 0.009 Pa·s and the experimental data from [53]. In addition, the simulation results using $\eta^*$ = 0.006 Pa·s and $\eta^*$ = 0.003 Pa·s are consistent with the empirical correlations derived by Bracke et al. [7] and Jiang et al [6], respectively. Note that instead of fitting the experimental data using two-parameter power law correlations, our prediction only depends on the value of interfacial viscosity $\eta^*$, which has a physical meaning as shown in Eq. (18) and can be identified independently from measurements of the apparent slip length. This verification on our simulation results demonstrate that not only the model can reproduce the rate-dependent behaviour of moving contact line, but also a good agreement with existing experimental studies can be achieved.

All the fitting results for $A$ and $B$, amplitude and exponent, respectively, under different magnitude of $\eta^*$ are plotted in Figure 12. In general, the value of $A$ and $B$ increases with $\eta^*$. The only exception is the dynamic receding angle with $\eta^*$ = 0.006 Pa·s. The reason is also due to water film formation caused by the strong adhesion between liquid and solid particles, as discussed in Figure 9. Therefore, the receding case with $\eta^* > 0.006$ Pa·s is not considered for results analysis. It can be concluded from our simulation that the value of power law fitting constants, $A$ and $B$, shares a positive correlation with $\eta^*$. Note for any given $\eta^*$, predictions can be made for both advancing and receding cases.

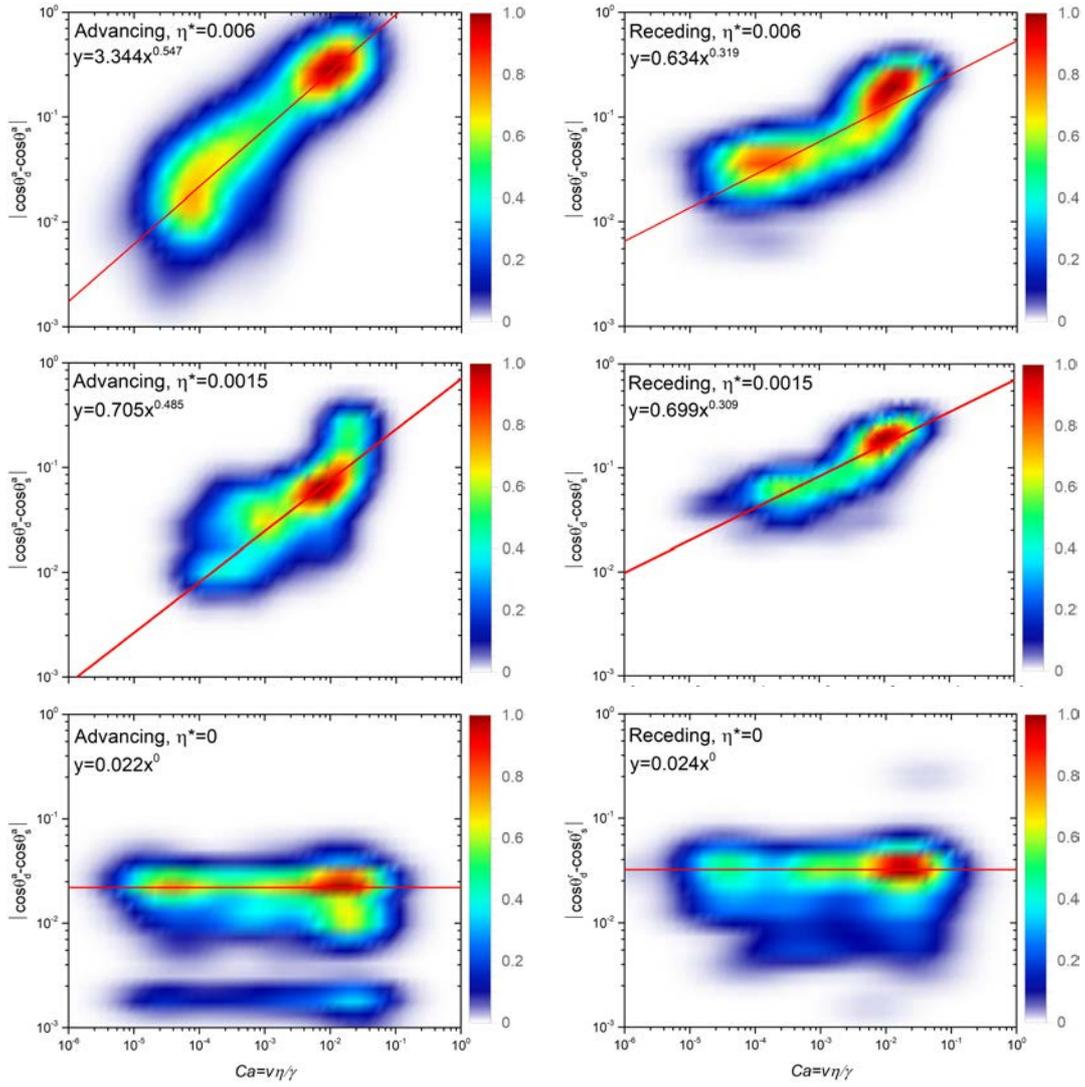

Fig. 10 Density histogram plots and power law fitting for dynamic contact angle results with different $\eta^*$ value ($\eta^*$=0.006, 0.0015 and 0 Pa·s), and x and y stands for $\left|\cos\theta_s^{a/r} - \cos\theta_d^{a/r}\right|$ and $Ca$, respectively.

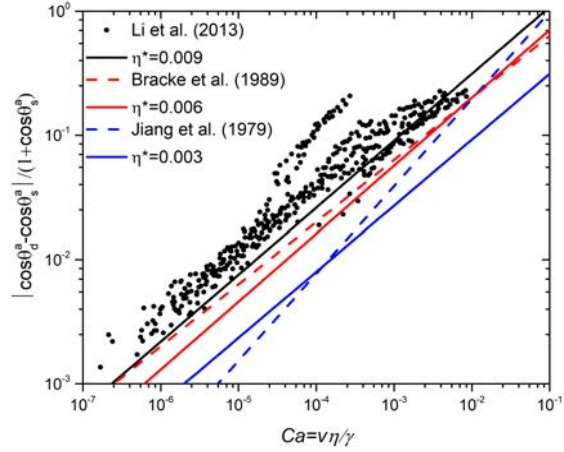

Fig. 11 Comparison of power law correlations obtained from simulations with experimental data and empirical power law correlations.

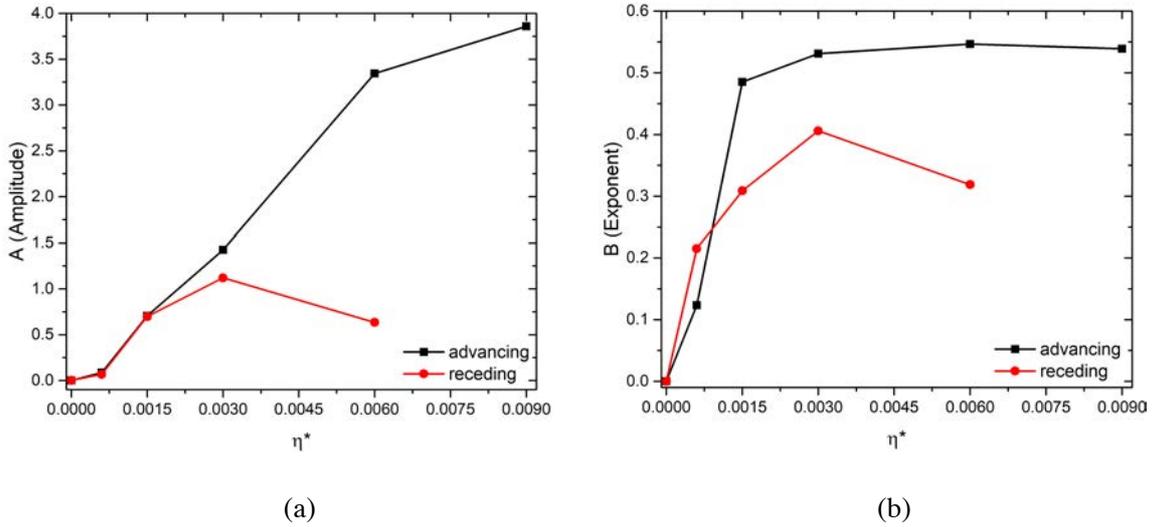

(a)                      (b)

Fig. 12 Relationship between $\eta^*$ and power law fitting constants: (a) amplitude and (b) exponent.

As stated in Section 3.2, to reproduce the dynamic contact angle, the viscosity $\eta$ at the liquid-solid interface is replaced by the interfacial viscous force parameter $\eta^*$, which is related to the slip length and surface roughness at the microscopic scale. The above-mentioned simulations are conducted with the same fluid viscosity and different power law fitting parameters are derived depending on the values of $\eta^*$. However, for fluids with different viscosity, the prediction and dependence of dynamic contact angle results on different value of $\eta^*$ is unknown. From Eq. (18), it is seen that the ratio of interfacial viscous force parameter $\eta^*$ and bulk viscosity $\eta$, i.e., $\eta^*/\eta$, dominates the power law correlation between dynamic contact angle and $Ca$, and same value of $\eta^*/\eta$ will reproduce similar results for fluids with different viscosity. To examine the influence of $\eta^*/\eta$ on dynamic contact angle simulation results, two additional sets of simulations with different bulk viscosity setting (0.013 Pa·s and 0.185 Pa·s) are conducted, and the results are compared with the $\eta^* = 0.003$ Pa·s and $\eta = 0.04$ Pa·s case discussed in Section 5.3. All these three cases have the same value of $\eta^*/\eta = 0.075$.

The data cover the range of $Ca$ from $10^{-6}$ to $10^{-1}$, and the results are plotted in Figure 12. The red line is the fitting function derived from the reference case with $\eta = 0.04$ Pa·s and $\eta^* = 0.003$ Pa·s. The dynamic receding case with fluid bulk viscosity 0.185 Pa·s is excluded for discussion, due to the fact that in this high viscosity setting the interfacial viscous force $\boldsymbol{F}_i^{vis}$ will be quite large with fixed ratio $\eta^*/\eta$, causing the strong adhesion between liquid and solid particles discussed previously and the determination of accurate contact angle becomes difficult.

In Figure 13, it is observed that the scattered data of three cases cover different range of $Ca$ as the viscosity is varied. Nonetheless, the data for advancing and receding cases with different bulk viscosity settings can still be described by the reference curve. This result demonstrates that for fluids with different viscosity, the same $\eta^*/\eta$ ratio will lead to similar dynamic contact angle as well as the corresponding power law fitting results.

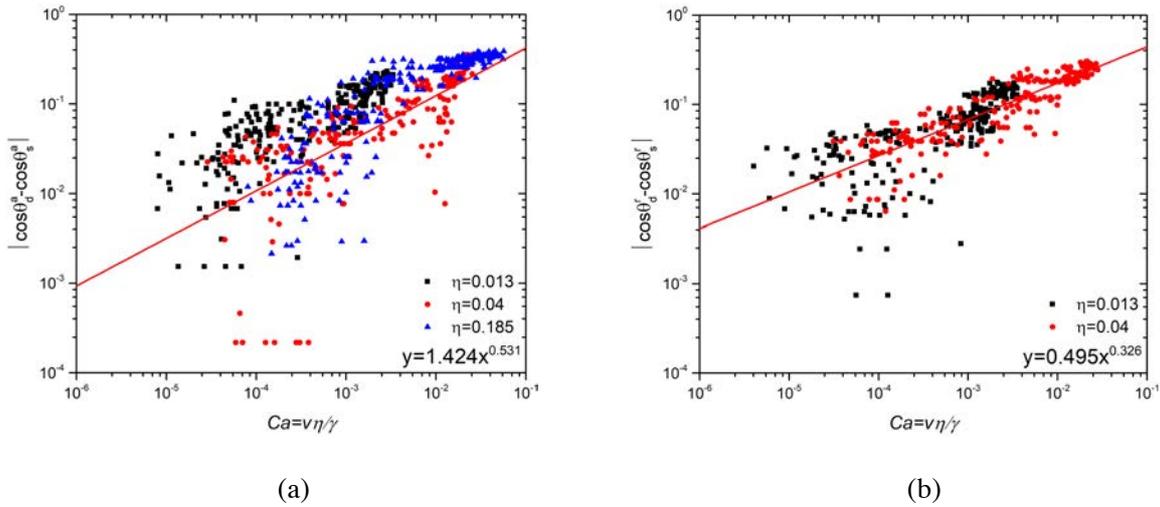

(a)          (b)

Fig. 13 Scattered plot of dynamic contact angles with different viscosity of fluid and $\frac{\eta^*}{\eta} = 0.075$: (a) advancing angle; (b) receding angle.

## 4. Conclusion

In this study, a modified SPH model with newly-introduced interfacial viscous force formulation is presented to simulate rate-dependent behaviour of moving contact line. The dynamic contact angle has been successfully reproduced with the implementation of interfacial viscous force. Correlations between simulated dynamic contact angle and $Ca$ are examined with empirical power law functions, and the results are in good agreement with experimental findings and theoretical analyses. Furthermore, the parametric study demonstrates the dependence of contact angle hysteresis and power law fitting on the magnitude of interfacial viscosity, which can be further related to measurable physical quantities, e.g., the slip length and microscopic surface roughness. This modified SPH model provides a simple and robust numerical solution to problems involving dynamic contact angle hysteresis, and the derived results can be further applied to a variety of industrial and geological applications where dynamic capillary interactions play a key role.

## Acknowledgements

This work was supported by Australian Research Council (Projects DP170102886) and The University of Sydney SOAR Fellowship. This research was undertaken with the assistance of the HPC service at The University of Sydney.

## Conflict of Interest Statement

Funding: This study was funded by Australian Research Council (grant number DP170102886).

Conflict of Interest: The authors declare that they have no conflict of interest.

## References

1. Abriola LM, Pinder GF (1985) A multiphase approach to the modeling of porous media contamination by organic compounds: 1. Equation development. Water resources research 21 (1):11-18
2. Holmes DW, Williams JR, Tilke P, Leonardi CR (2016) Characterizing flow in oil reservoir rock using SPH: absolute permeability. Computational Particle Mechanics 3 (2):141-154
3. Bandara UC, Palmer BJ, Tartakovsky AM (2016) Effect of wettability alteration on long-term behavior of fluids in subsurface. Computational Particle Mechanics 3 (2):277-289
4. Kiwi-Minsker L, Renken A (2005) Microstructured reactors for catalytic reactions. Catalysis today 110 (1):2-14
5. Schwartz AM, Tejada SB (1972) Studies of dynamic contact angles on solids. Journal of Colloid and Interface Science 38 (2):359-375
6. Jiang T-S, Soo-Gun O, Slattery JC (1979) Correlation for dynamic contact angle. Journal of Colloid and Interface Science 69 (1):74-77
7. Bracke M, De Voeght F, Joos P (1989) The kinetics of wetting: the dynamic contact angle. Trends in Colloid and Interface Science III:142-149
8. Cox R (1986) The dynamics of the spreading of liquids on a solid surface. Part 1. Viscous flow. Journal of Fluid Mechanics 168:169-194
9. Hoffman RL (1983) A study of the advancing interface: II. Theoretical prediction of the dynamic contact angle in liquid-gas systems. Journal of Colloid and Interface Science 94 (2):470-486

10. Raiskinmäki P, Shakib-Manesh A, Jäsberg A, Koponen A, Merikoski J, Timonen J (2002) Lattice-Boltzmann simulation of capillary rise dynamics. Journal of statistical physics 107 (1):143-158
11. Koplik J, Banavar JR, Willemsen JF (1988) Molecular dynamics of Poiseuille flow and moving contact lines. Physical review letters 60 (13):1282
12. Huber M, Keller F, Säckel W, Hirschler M, Kunz P, Hassanizadeh SM, Nieken U (2016) On the physically based modeling of surface tension and moving contact lines with dynamic contact angles on the continuum scale. Journal of Computational Physics 310:459-477
13. Lukyanov AV, Likhtman AE (2016) Dynamic Contact Angle at the Nanoscale: A Unified View. ACS nano 10 (6):6045-6053
14. Tartakovsky AM, Panchenko A (2016) Pairwise force smoothed particle hydrodynamics model for multiphase flow: surface tension and contact line dynamics. Journal of Computational Physics 305:1119-1146
15. Eral H, Oh J (2013) Contact angle hysteresis: a review of fundamentals and applications. Colloid and polymer science 291 (2):247-260
16. Rame E (1997) The interpretation of dynamic contact angles measured by the Wilhelmy plate method. Journal of colloid and interface science 185 (1):245-251
17. Blake T, Haynes J (1969) Kinetics of liquidliquid displacement. Journal of colloid and interface science 30 (3):421-423
18. Petrov P, Petrov I (1992) A combined molecular-hydrodynamic approach to wetting kinetics. Langmuir 8 (7):1762-1767
19. Elliott G, Riddiford A (1967) Dynamic contact angles: I. The effect of impressed motion. Journal of colloid and interface science 23 (3):389-398
20. Schäffer E, Wong P-z (2000) Contact line dynamics near the pinning threshold: A capillary rise and fall experiment. Physical Review E 61 (5):5257
21. Shi Z, Zhang Y, Liu M, Hanaor DA, Gan Y (2017) Dynamic contact angle hysteresis in liquid bridges. arXiv preprint arXiv:171204703
22. Kim J-H, Rothstein JP (2015) Dynamic contact angle measurements of viscoelastic fluids. Journal of Non-Newtonian Fluid Mechanics 225:54-61
23. Seebergh JE, Berg JC (1992) Dynamic wetting in the low capillary number regime. Chemical Engineering Science 47 (17-18):4455-4464
24. Meakin P, Tartakovsky AM (2009) Modeling and simulation of pore- scale multiphase fluid flow and reactive transport in fractured and porous media. Reviews of Geophysics 47 (3)
25. Thompson PA, Robbins MO (1989) Simulations of contact-line motion: slip and the dynamic contact angle. Physical Review Letters 63 (7):766
26. Dos Santos LO, Wolf FG, Philippi PC (2005) Dynamics of interface displacement in capillary flow. Journal of statistical physics 121 (1-2):197-207
27. Chibbaro S, Biferale L, Diotallevi F, Succi S (2009) Capillary filling for multicomponent fluid using the pseudo-potential Lattice Boltzmann method. The European Physical Journal-Special Topics 171 (1):223-228
28. Caiazzo A (2008) Analysis of lattice Boltzmann nodes initialisation in moving boundary problems. Progress in Computational Fluid Dynamics, an International Journal 8 (1-4):3-10
29. Becker M, Teschner M Weakly compressible SPH for free surface flows. In: Proceedings of the 2007 ACM SIGGRAPH/Eurographics symposium on Computer animation, 2007. Eurographics Association, pp 209-217
30. Monaghan JJ (1994) Simulating free surface flows with SPH. Journal of computational physics 110 (2):399-406
31. Breinlinger T, Polfer P, Hashibon A, Kraft T (2013) Surface tension and wetting effects with smoothed particle hydrodynamics. Journal of Computational Physics 243:14-27

32. Monaghan JJ (1992) Smoothed particle hydrodynamics. Annual review of astronomy and astrophysics 30 (1):543-574
33. Liu MB, Liu GR (2010) Smoothed Particle Hydrodynamics (SPH): an Overview and Recent Developments. Archives Of Computational Methods In Engineering 17 (1):25-76. doi:10.1007/s11831-010-9040-7
34. Meister M, Burger G, Rauch W (2014) On the Reynolds number sensitivity of smoothed particle hydrodynamics. Journal of Hydraulic Research 52 (6):824-835
35. Morris JP, Fox PJ, Zhu Y (1997) Modeling low Reynolds number incompressible flows using SPH. Journal of computational physics 136 (1):214-226
36. Li L, Shen L, Nguyen GD, El-Zein A, Maggi F (2018) A smoothed particle hydrodynamics framework for modelling multiphase interactions at meso-scale. Computational Mechanics:1-15
37. Monaghan JJ, Kajtar JB (2009) SPH particle boundary forces for arbitrary boundaries. Computer Physics Communications 180 (10):1811-1820
38. Colagrossi A, Landrini M (2003) Numerical simulation of interfacial flows by smoothed particle hydrodynamics. Journal of Computational Physics 191 (2):448-475
39. Feldman J, Bonet J (2007) Dynamic refinement and boundary contact forces in SPH with applications in fluid flow problems. International Journal for Numerical Methods in Engineering 72 (3):295-324
40. Wang J, Chan D (2014) Frictional contact algorithms in SPH for the simulation of soil–structure interaction. International Journal for Numerical and Analytical Methods in Geomechanics 38 (7):747-770
41. Schnell E (1956) Slippage of water over nonwettable surfaces. Journal of Applied Physics 27 (10):1149-1152
42. Churaev N, Sobolev V, Somov A (1984) Slippage of liquids over lyophobic solid surfaces. Journal of Colloid and Interface Science 97 (2):574-581
43. Cheng J-T, Giordano N (2002) Fluid flow through nanometer-scale channels. Physical review E 65 (3):031206
44. Choi C-H, Westin KJA, Breuer KS (2003) Apparent slip flows in hydrophilic and hydrophobic microchannels. Physics of fluids 15 (10):2897-2902
45. Pit R, Hervet H, Leger L (2000) Direct experimental evidence of slip in hexadecane: solid interfaces. Physical review letters 85 (5):980
46. Rothstein JP (2010) Slip on superhydrophobic surfaces. Annual Review of Fluid Mechanics 42:89-109
47. Majumder M, Chopra N, Andrews R, Hinds BJ (2005) Nanoscale hydrodynamics: enhanced flow in carbon nanotubes. Nature 438 (7064):44-44
48. Lee C, Kim C-JC (2009) Maximizing the giant liquid slip on superhydrophobic microstructures by nanostructuring their sidewalls. Langmuir 25 (21):12812-12818
49. Ramachandran P PySPH: a reproducible and high-performance framework for smoothed particle hydrodynamics. In: Proceedings of the 15th Python in Science Conference, 2016. pp 127-135
50. Liu G-R, Liu MB (2003) Smoothed particle hydrodynamics: a meshfree particle method. World Scientific,
51. Niavarani A, Priezjev NV (2010) Modeling the combined effect of surface roughness and shear rate on slip flow of simple fluids. Physical Review E 81 (1):011606
52. Karim AM, Rothstein JP, Kavehpour HP (2018) Experimental study of dynamic contact angles on rough hydrophobic surfaces. Journal of colloid and interface science 513:658-665
53. Li X, Fan X, Askounis A, Wu K, Sefiane K, Koutsos V (2013) An experimental study on dynamic pore wettability. Chemical Engineering Science 104:988-997